\documentclass[preprint,showpacs
]{revtex4-1}
\usepackage{graphicx}
\usepackage{amsfonts}
\usepackage{amssymb}
\usepackage{amsthm}
\usepackage{array}
\usepackage{amsmath}
\usepackage{verbatim} 
\usepackage{hyperref}
\usepackage{color}
\usepackage{bbold}
\usepackage{epstopdf}
\usepackage{mathtools}
\usepackage{enumerate}
\usepackage{subcaption}
\usepackage{color}

\newtheorem*{proposition*}{Proposition}

\newtheorem*{theorem*}{Theorem}

\newtheorem*{corollary*}{Corollary}

\newcommand{\ket}[1]{\left\vert#1\right\rangle}
\newcommand{\bra}[1]{\left\langle#1\right\vert}

\newcommand{\cnot}[1]{U^\mathrm{CNOT}_{ #1 } }
\newcommand{\cz}[1]{U^\mathrm{CZ}_{ #1 } }

\def\bra#1{\langle #1|}
\def\ket#1{\left|#1 \right>}

\def\Tr{\mbox{Tr}}

\begin{document}
\title{Coherence as a Unit Resource for Quantum Error Correction}
\author{Kok Chuan Tan, S. Omkar, Hyeonseok Jeong}
\email{bbtankc@gmail.com}
\affiliation{Center for Macroscopic Quantum Control, Department of Physics and Astronomy, Seoul National University, Seoul, 151-742, Korea}
\date{\today}

\begin{abstract}
In this paper we study an error correcting protocol that specifically derives its error correcting properties from elementary units of coherence. The entire protocol from beginning to end is performed using non-coherence increasing operations, resulting in the consumption of the input coherence, thus necessitating further quantum resources if one wishes to perform the protocol again. We show that even when the input quantum resource is just 1 coherent qubit, one may acquire partial protection from phase flip errors, and that this can be scaled up to protect against arbitrary qubit errors with 6 ancillary coherent qubits as input. The work presented strengthens the operational interpretation of a single unit of coherence by providing a useful information theoretic task that one may perform when such elementary units of coherence are available.
\end{abstract}

\maketitle

\section{Introduction} 

Quantum information science has gained prominence as an area of research in the recent decades. One of the key promises of the field is that the the quantum regime has certain intrinsic advantages over  classical theories, advantages that can be exploited for a variety of informational tasks. As a result, there has been renewed interest in the study of the differences between quantum and classical theories. One of the key insights that has emerged from such studies is that many of these quantum advantages can be classified and directly attributed to specific quantum resources, and that such quantum resources can be quantified in a natural way via what are now collectively called \textit{quantum resource theories}.

A recently developed approach towards the quantification of quantum resources comes from the so called \textit{resource theory of coherence}, first formalized by Baumgratz \textit{et al.}~\cite{Baumgratz14}. This resource theory of coherence essentially lays out certain ground rules in order to quantify the nonclassicality that can be attributed to the off diagonal elements of the quantum state. Under this approach, a set of states and a set of quantum operations are identified to be ``free". The former because such states are considered classical, and the latter because it doesn't introduce nonclassical resources into the system of interest. Correspondingly, such free states and operations are referred to as incoherent states and incoherent operations. In another important recent development, it is shown that if one has many copies of a given quantum states that exhibit coherence in the previously mentioned sense, then this coherence may be distilled into pure qubit states that exhibit maximal coherence using only incoherent operations~\cite{Winter2016}. In this way, the amount of coherence in a quantum system can then be attributed operational significance, in terms of the number of pure maximally coherent qubits one can distil from the state. Since its initial proposal, this particular notion of quantum coherence has been applied to study a variety of physical phenomena. Some examples include topics as diverse as quantum correlations~\cite{Tan2016, Streltsov15}, biological systems, macroscopicity~\cite{Yadin2015, Kwon2016}, quantum optics~\cite{Bagan2016, Tan2017} and quantum metrology~\cite{Giorda2016}.

While the basic tools for the quantification of quantum coherences are quite robust at this point, much less is known about how to best exploit this nonclassical feature for informational tasks. On this front, some recent results have shown that quantum coherence plays a role in the accuracy/precision of some quantum algorithms~\cite{Hillery2016,Matera2016}. In this paper, we consider another useful informational task: quantum error correction. Since the initial conception of a quantum computer, some form of quantum error correction was immediately recognized as necessary in order to realize a fully functioning quantum computer. Shor was the first to demonstrate that quantum computing can be made fault tolerant by designing the first quantum error correcting procedure~\cite{Shor1996}. Since this seminal work, many more error correcting protocols have been proposed~\cite{Terhal2015} and subsequently implemented in the laboratory~\cite{Chiaverini2004, Reed2012}.

For this paper, we are primarily interested in the resource theoretical aspects of error correction. In the subsequent sections, we detail an explicit error correcting protocol where the basic unit of coherence, a maximally coherent qubit, is used as the fundamental resource. The error correction requires only 2 basic ingredients: 1) ancillary qubits that are initiallized in the maximally coherent state and 2) incoherent unitary operations, exactly as identified in \cite{Baumgratz14}. This is sufficient to establish that coherence is the primary resource exploited in the procedure. At the end of every error correcting procedure, incoherent measurements will destroy the initial coherence used to perform the logical encoding, thus necessitating additional coherent qubits as input if the protocol were to be performed again. In this way, a single maximally coherent qubit may be associated with a single unit of quantum error correcting property. This further strengthens the results by Winter and Yang~\cite{Winter2016} by demonstrating a useful informational task that can be performed using distilled, maximally coherent qubits. This also simultaneously demonstrates that every quantum state that is coherent with respect to the computational basis is useful for quantum error correction, since there exist no "bound coherence", so maximally coherent qubits can always be distilled from coherent states.

\section{Preliminaries}

The notion of coherence that we will employ in this paper will be the one identified in~\cite{Baumgratz14}, where a set of axioms is specified for quantifiers of quantum coherence. For the convenience of the reader, we replicate these axioms below:

For a fixed basis $\{ \ket{i} \}$, the set of incoherent states $\cal I$ is the set of quantum states with diagonal density matrices with respect to this basis. An incoherent completely positive and trace preserving maps (ICPTP) is one that maps every incoherent state to another incoherent state. Given this, we say that  $\mathcal{C}$ is a measure of quantum coherence if it satisfies following properties:
(C1) $\mathcal{C}(\rho) \geq 0$ for any quantum state $\rho$ and equality holds if and only if $\rho \in \cal I$.
(C2a) The measure is non-increasing under a ICPTP map $\Phi$ , i.e., $C(\rho) \geq C(\Phi(\rho))$.
(C2b) Monotonicity for average coherence under selective outcomes of ICPTP:
$C(\rho) \geq \sum_n p_n C(\rho_n)$, where $\rho_n = K_n \rho K_n^\dagger/p_n$ and $p_n = \Tr [K_n \rho K^\dagger_n ]$ for all $K_n$ with $\sum_n K_n K^\dagger_n = \mathbb 1$ and $K_n {\cal I} K_n^\dagger \subseteq \cal I$.
(C3) Convexity, i.e. $\lambda C(\rho) + (1-\lambda) C(\sigma) \geq C(\lambda \rho + (1-\lambda) \sigma)$, for any density matrix $\rho$ and $\sigma$ with $0\leq \lambda \leq 1$.

It turns out that the optimal rate at which you can distil maximally coherent qubits in the infinite copy limit satisfies all the above axioms~\cite{Winter2016}. Relevant to our interests are axioms (C2a) and (C2b), which specifies that incoherent operations (i.e. ICPTP maps) cannot increase the amount of coherence within the system. One may check that a particular operation is incoherent if it always maps a diagonal density matrix to another diagonal density matrix. One important example of such an operation is the CNOT gate. It is clear that its action on classical bits is simply a classical CNOT operation, so it is an incoherent operation. In contrast, a CNOT operation does not fall under the regime of local operations and classical communication~\cite{Horodecki2001}, which form a set of non entanglement increasing operations. For some given set of basis states $\{ \ket{i} \}_{i = 1,\ldots, d}$, all ICPTP maps $\Phi_{\mathrm{ICPTP}}$ are prescribed by some set of Kraus operators $\{ K_l \}$ \cite{kraus} of the form $K_j = \sum_i c(i)\ket{f_j(i)} \bra{i}$ such that $\sum_l K_l^\dag K_l = \openone$, $\Phi_{\mathrm{ICPTP}}(\rho) = \sum_l  K_l \rho K_l^\dag$ and $f_j$ is some function on integers.

We now introduce some conventions that will be used in the subsequent sections. We will denote the canonical Pauli matrices on the $m$th qubit as $X_m$, $Y_m$ and $Z_m$ respectively. The computational basis refers to the basis $\{ \ket{0}, \ket{1} \}$, from which we can define the states $\ket{+} \coloneqq  \frac{1}{\sqrt{2}}(\ket{0}+ \ket{1})$ and $\ket{-} \coloneqq  \frac{1}{\sqrt{2}}(\ket{0} - \ket{1})$. The unitary performing a CNOT operation between the $m$th and $n$th qubits is denoted $U^{\mathrm{CNOT}}_{mn}$ where the first subindex $m$ is the control qubit, i.e. $U^{\mathrm{CNOT}}_{mn}\ket{0}_m \ket{\psi}_n = \ket{0}_m \ket{\psi}_n$ and $U^{\mathrm{CNOT}}_{mn}\ket{1}_m \ket{\psi}_n = \ket{1}_m X_n\ket{\psi}_n$. In a similar manner, we will denote a controlled phase flip, also called the CZ operation, on the $\{ \ket{+}, \ket{-} \}$ basis as $U^{\mathrm{CZ}}_{mn}$. The controlled phase flip performs the operation $U^{\mathrm{CZ}}_{mn}\ket{+}_m \ket{\psi}_n = \ket{+}_m \ket{\psi}_n$ and $U^{\mathrm{CZ}}_{mn}\ket{-}_m \ket{\psi}_n = \ket{-}_m Z_n \ket{\psi}_n$. We note that this final definition is a slight deviation from the usual convention, where the control operation is controlled on the computational basis $\{ \ket{0}, \ket{1} \}$. As previously discussed, the quantum CNOT operation operating on diagonal density matrices is just a classical CNOT operation, so $\cnot{12}\ket{i}_1\ket{j}_2 = \ket{i}_1\ket{j \oplus i}_2 $, and thus is an incoherent operation. It can be further verified that $\cz{mn} \ket{i}_m\ket{j}_n = \cnot{nm}\ket{i}_m\ket{j}_n = \ket{i \oplus j}_m\ket{j}_n$ so it is also an incoherent. Furthermore, the single qubit Pauli operators $X_m$, $Y_m$ and $Z_m$ are also incoherent. These operations will form the basis for the subsequent error correcting procedure.

\section{Partial Error Correction}

In order to illustrate the process we begin with the following observations, which will be used as a primitive to build the complete code.

Consider the states $\ket{+}$ and $\ket{-}$. These are both maximally coherent qubits, but we will choose the reference state to be $\ket{+}$. Consider now the product state $\ket{+}_1\ket{0}_2$. The CNOT operation $\cnot{1,2}$ with qubit 1 as the control qubit will the maximally entangled state $\ket{\Phi^+}_{12} = \frac{1}{\sqrt{2}} (\ket{00}_{12} + \ket{11}_{12})$ after acting on the product state. Suppose we perform a phase flip operation $Z_2$ on qubit 2. It is then straightforward to see that the the resulting state is now  $\ket{\Phi^-}_{12} = \frac{1}{\sqrt{2}} (\ket{00}_{12} - \ket{11}_{12})$. The application of another CNOT operation will result in $\cnot{12}\ket{\Phi^-} = \ket{-}\ket{0}$, where we can clearly see that the error is propagated from the second qubit to the first, with the transformation of $\ket{+}_1$ to $\ket{-}_1$ indicating a phase flip error has occurred on the 2nd qubit. As such, a single maximally coherent qubit, together with the CNOT operation, confers some partial error correcting property to some qubit of interest. Since this partial error correcting property comes from the combination of a maximally coherent qubit together with an incoherent operation, we can surmise that the protection is derived from coherence.

To complete the discussion, let us also consider the case when the initial state is the product state $\ket{+}_1\ket{1}_2$. Applying again the CNOT operations, the resulting state is $\ket{\Psi^+} = \frac{1}{\sqrt{2}}(\ket{01}_{12}+\ket{10}_{12})$. If a phase flip error happens on the second bit, the resulting state is $- \ket{\Psi^-} = \frac{1}{\sqrt{2}}(\ket{01}_{12}-\ket{10}_{12})$. Applying the final CNOT operation, we retrieve $-\ket{-}_1\ket{1}_2$, so again, a phase flip on qubit 2 will be propagated to qubit 1, transforming   $\ket{+}_1$ to $\ket{-}_1$. The additional global phase can be corrected if desired since qubit 1 contains the syndrome outcome by applying a controlled $Z$ operation  which we define to be  $\cz{12} \ket{+}_1\ket{i}_2 = \ket{+}_1\ket{i}_2$ and $\cz{12} \ket{-}_1\ket{i}_2 = \ket{-}_1Z_2\ket{i}_2$ where $i=0,1$. This gives us $\cz{12}\cnot{12}\ket{\Psi-} = \ket{-}_1 \ket{1}_2$. We note also that $\cz{12}\cnot{12}\ket{\Phi-} = \ket{-}_1 \ket{0}_2$, so the decoding procedure $\cz{12}\cnot{12}$ will properly correct a phase flip error in both cases. Since $\cz{12}$ is also an incoherent operation, qubit 2 is completely protected from phase flip errors.

Suppose now that we have a superposition of errors of the form $\mathcal{E} = a \openone_2 + b Z_2$, and some arbitrary initial state of the form $\ket{\psi} = \alpha \ket{0} + \beta \ket{1}$. Using the above arguments, we can protect qubit 2 from such an error without performing a syndrome measurements, using only the incoherent, CNOT and CZ operations. After the first CNOT, we will have the state $\cnot{12}\ket{+}\ket{\psi} = \alpha \ket{\Phi^+} + \beta \ket{\Psi^+}$. After the error $\mathcal{E}$, the resulting state is $\mathcal{E}_2\cnot{12}\ket{+}\ket{\psi} = a(\alpha \ket{\Phi^+} + \beta \ket{\Psi^+})+ b(\alpha \ket{\Phi^-} + \beta \ket{\Psi^-})$. Finally, we decode by applying  CNOT again and a final controlled Z operation to correct the phase flip, resulting in the state $\cz{12} \cnot{12}\mathcal{E}_2\cnot{12}\ket{+}\ket{\psi} = a(\alpha \ket{+}_1 \ket{0}_2 + \beta \ket{+}_1 \ket{1}_2) + b(\alpha \ket{-}_1 \ket{0}_2 + \beta \ket{-}_1 \ket{1}_2) = (a\ket{+}_1+b\ket{-}_1) \alpha \ket{0}_2 + (a\ket{+}_1+b\ket{-}_1) \beta \ket{1}_2 = (a\ket{+}_1+b\ket{-}_1) \ket{\psi}_2$. After the decoding process, we see that $\ket{\psi}_2$ remains unperturbed while qubit 1 has completely absorbed the error, thus correcting for the error on qubit 2.

Unfortunately, while the above encoding and decoding processes work for correcting phase flip errors on qubit 2, the situation becomes more complex when we consider errors on qubit 1, which is necessary to achieve full error correction. The main complication comes from the fact that $Z_1 \cnot{12}\ket{+}_1\ket{1}_2 = Z_1 \ket{\Psi^+}_{12} = - Z_2 \ket{\Psi^+}_{12}$, so a phase flip error on the first qubit will result in an additional negative phase. The decoding process described above is unable to detect the presence of this additional phase, which inevitably implies that when a phase flip occurs on qubit 1, the initial state $\ket{\psi}_2 = \alpha \ket{0}_2 + \beta \ket{1}_2$ will be transformed to $\ket{\psi'}_2 = \alpha \ket{0}_2 - \beta \ket{1}_2$ after decoding, i.e. a phase flip error on qubit 1 is propagated to qubit 2, so the process works both ways.

However, as we shall see, complete error correction can be achieved my simply adding more  coherent qubits to the the encoding process. This is intuitive from the resource perspective, since it suggests that the introduction of more nonclassical resources confers additional error correcting benefits. As we shall see, the correction of a single $Z$ error will require 2 additional maximally coherent qubits, and the full correction of $X,Y, Z$ errors will require 6 additional maximally coherent qubits, loosely corresponding to 
 2 additional maximally coherent qubits for each error that can occur.

\section{Correcting Phase Flips}

With the previous discussion, we now have the tools to construct the full error correcting code. The procedure will be laid out in full in this section and the next. 

We first scale up the code to protect against random phase flips. The basic premise is just as before, but we now introduce an additional $\ket{+}$ qubit. Let us define the logical qubits $$\ket{0_l}_{123} =  \cnot{32}\cnot{12}\ket{+}_1 \ket{0}_2 \ket{+}_3 = \ket{000}_{123}+\ket{011}_{123}+\ket{101}_{123} +\ket{110}_{123}$$ and $$\ket{1_l}_{123} =  \cnot{32}\cnot{12}\ket{+}_1 \ket{1}_2 \ket{+}_3 = \ket{111}_{123}+\ket{100}_{123}+\ket{010}_{123}+\ket{001}_{123}$$

where we have omitted the normalization factor. The normalization factor will be omitted from this point onwards unless required. Note that both logical qubits are symmetric with respect to qubit swap operations between any 2 qubits. 

A manual calculation will lead to the following error table:

\begin{center}
\begin{tabular}{|c|c|c|}
\hline
Error $\mathcal{E}$ & $\cnot{12}\cnot{32}\mathcal{E}\ket{0_l}_{123}$ & $\cnot{12}\cnot{32}\mathcal{E}\ket{1_l}_{123}$ \\ \hline
$\openone$ & $\ket{+}_1 \ket{0}_2 \ket{+}_3$ &  $\ket{+}_1 \ket{1}_2 \ket{+}_3$ \\ \hline
$Z_1$ & $\ket{-}_1 \ket{0}_2 \ket{+}_3$ &  $\ket{-}_1 \ket{1}_2 \ket{+}_3$ \\ \hline
$Z_2$ & $\ket{-}_1 \ket{0}_2 \ket{-}_3$ &  $-\ket{-}_1 \ket{1}_2 \ket{-}_3$ \\ \hline
$Z_3$ & $\ket{+}_1 \ket{0}_2 \ket{-}_3$ &  $\ket{+}_1 \ket{1}_2 \ket{-}_3$ \\ 
\hline

\end{tabular}
\end{center}

From the above table, we see that every error is correctly propagated to qubits 1 and 3, regardless of where they occur. It can be shown that the negative phase can be corrected via an incoherent measurement process. Consider the Kraus maps $K^0 = \ket{0}\bra{+}$ and $K^1 = \ket{1}\bra{-}$ corresponding to measurement outcomes 0 and 1 respectively. It is readily seen that it is a complete set of Kraus operators since $(K^0)^\dag K^0+(K^1)^\dag K^1 = \openone$. Furthermore, it always maps a state in the computational basis to white noise, i.e. $\mathcal{M}(\ket{i}\bra{i}) \coloneqq \sum_{j=0}^1 K^j \ket{i}\bra{i} (K^j)^\dag = \frac{1}{2}\openone$ so it is an incoherent operation. After performing this incoherent measurement, the information in qubits 1 and 3 can be read out from the computational basis, allowing us to correct for the phase by performing a phase flip operation $ \ket{1}_1 Z_2 \ket{i}_2 \ket{1}_3$ when the measurements on both qubits 1 and 3 returns the outcome 1. The error is thus corrected incoherently, consuming the input coherence in the process since the states $\ket{+}, \ket{-}$ are mapped onto $\ket{0}, \ket{1}$ respectively.

Note that the additional negative phase can also be alternatively corrected by a conditional phase flip $\cz{(13)2}$ which performs the necessary controlled phase flip operation in the case $\cz{(13)2}\ket{-}_1 \ket{i}_2 \ket{-}_3 = \ket{-}_1 Z_2\ket{i}_2 \ket{-}_3$ and does nothing (or the identity operation) otherwise. However, $\cz{(13)2}$ is not an incoherent operation. This can be demonstrated via the following counter example. Consider the incoherent pure state $\ket{0}_1\ket{1}_2\ket{0}_3 = \ket{1}_2(\ket{+}_1\ket{+}_3+\ket{+}_1\ket{-}_3+\ket{-}_1\ket{+}_3+\ket{-}_1\ket{-}_3)$. It can be directly verified that $\cz{(13)2} \ket{0}_1\ket{1}_2\ket{0}_3  =  \ket{0}_1\ket{1}_2\ket{+}_3 + \ket{1}_1\ket{1}_2\ket{-}_3$, which is an entangled coherent state. $\cz{(13)2}$ is therefore not an incoherent operation. However, the fact that the protocol can be performed using only incoherent operations suggest that the error correcting property is derived from coherence. We also note that 3 qubit codes of a similar type have already been performed in the laboratory~\cite{Chiaverini2004}.

\section{Correcting Arbitrary Single Qubit Errors } 

Full correction of an arbitrary single qubit error will require yet another layer of encoding. First, we observe that the above encoding allows for the convenient property that $X_m \ket{0_l}_{123} = \ket{1_l}_{123}$ where $m = 1,2,3$. This implies that a bit flip on the individual qubit level is equivalent to a bit flip on the logical level. This makes correction of any single bit flip a relatively easy affair, since a phase flip is simply a rotated bit flip on the level of the Bloch sphere. 

We now introduce some new notations. Let us define $\ket{i^{(jk)}}_{123} \coloneqq \cnot{32}\cnot{12} \ket{j}_1 \ket{i}_2 \ket{k}_3 $, where $i = 0,1$ and $j,k = +,-$. So, for instance, from our previous definition we have $\ket{0_l}_{123} =  \cnot{32}\cnot{12}\ket{+}_1 \ket{0}_2 \ket{+}_3 = \ket{0^{(++)}}_{123} $. From this, we can further define 3 qubit logical operations that affect only the logical degrees of freedom, while leaving the superscript intact. For instance, a 3 qubit bit flip is defined to be $X_{(123)} \coloneqq \cnot{32}\cnot{12} X_2 \cnot{12}\cnot{32}$. It can be verified that $X_{(123)} \ket{i^{(j,k)}}_{123} = \cnot{32}\cnot{12} \ket{j}_1 X_2\ket{i}_2 \ket{k}_3 = \ket{(i \oplus 1) ^{(j,k)}}_{123} $ and  a similar definition for the phase flip gives $Z_{(123)} \ket{i^{(j,k)}}_{123} = (-1)^i\ket{i ^{(j,k)}}_{123}$ so that indeed, these operations operate only the the logical degrees of freedom encoded by $i$. We can also define operations between 2, 3-qubit clusters, such as the CZ operation between qubit clusters denoted by $a = a_1a_2a_3$ and $b = b_1b_2b_3$. In this case, the operator is defined to be $\cz{ab} \coloneqq \cnot{a_3a_2}\cnot{a_1a_2} \cnot{b_3b_2}\cnot{b_1b_2} \cz{a_2b_2} \cnot{a_1a_2}\cnot{a_3a_2} \cnot{b_1b_2}\cnot{b_3b_2}$. As before, it can be verified that $\cnot{ab}\ket{i^{(jk)}}_a\ket{l^{(mn)}}_b = (\cnot{a_3a_2}\cnot{a_1a_2} \cnot{b_3b_2}\cnot{b_1b_2}) \cz{a_2b_2}\ket{j}_{a_1} \ket{i}_{a_2} \ket{k}_{a_3}\ket{m}_{b_1} \ket{l}_{b_2} \ket{n}_{b_3}$, which again, only operates on the logical degrees of freedom indicated by $i,l$. Crucially, such operations are incoherent operations as they are composed of only 2 qubit CNOT and CZ operations, both of which are already demonstrated to be incoherent.

Based on this notation, the full logical encoding that corrects both bit and phase flips is the following:

$$
\ket{0_L}_{abc} = \cz{cb}\cz{ab}\ket{0^{(++)}}_a \ket{0^{(++)}}_b \ket{0^{(++)}}_c = \ket{0^{(++)}}_a \ket{0^{(++)}}_b \ket{0^{(++)}}_c
$$

and

$$
\ket{1_L}_{abc} = \cz{cb} \cz{ab}\ket{0^{(++)}}_a \ket{1^{(++)}}_b \ket{0^{(++)}}_c = \ket{1^{(++)}}_a \ket{1^{(++)}}_b \ket{1^{(++)}}_c
$$

where $a=a_1a_2a_3$, $b= b_1b_2b_3$ and $c= c_1c_2c_3$. The controlled Z type operation explicit indicates that the correction of bit flips employs the same encoding as for phase flips, but on a rotated basis. A bit flip on any single qubit will give rise to the following error table:

\begin{center}
\begin{tabular}{|c|c|c|}
\hline
Error $\mathcal{E}$ & $\cz{ab}\cz{cb}\mathcal{E}\ket{+_L}_{abc}$ & $\cz{ab}\cz{cb}\mathcal{E}\ket{-_L}_{abc}$ \\ \hline
$\openone$ & $\ket{0^{(++)}}_a \ket{+^{(++)}}_b \ket{0^{(++)}}_c$ &  $\ket{0^{(++)}}_a \ket{-^{(++)}}_b \ket{0^{(++)}}_c$ \\ \hline
$X_{a_i}$ & $\ket{1^{(++)}}_a \ket{+^{(++)}}_b \ket{0^{(++)}}_c$ &  $\ket{1^{(++)}}_a \ket{-^{(++)}}_b \ket{0^{(++)}}_c$ \\ \hline
$X_{b_i}$ & $\ket{1^{(++)}}_a \ket{+^{(++)}}_b \ket{1^{(++)}}_c$ &  $-\ket{1^{(++)}}_a \ket{-^{(++)}}_b \ket{1^{(++)}}_c$ \\ \hline
$X_{c_i}$ & $\ket{0^{(++)}}_a \ket{+^{(++)}}_b \ket{1^{(++)}}_c$ &  $\ket{0^{(++)}}_a \ket{-^{(++)}}_b \ket{1^{(++)}}_c$ \\ 
\hline

\end{tabular}
\end{center}

which is similar as for the phase flip error but in a rotated basis. The additional negative phase can therefore be corrected in a similar manner, this time by performing a bit flip $X_{b_2}$ if a classical measurement on qubits $a_2$ and $c_2$ corresponding to projectors $\ket{0}\bra{0}$ and $\ket{1}\bra{1}$ returns the outcome 1. Any single qubit phase flip will affect the superscript $(++)$ according to the error table for the phase flip, and the error can be corrected by performing a phase flip $Z_{b_2}$ when the outcomes of the incoherent measurement given by $K^0 = \ket{0}\bra{+}$ and $K^1 = \ket{1}\bra{-}$ performed on the qubit pairs $(a_1,a_2)$, $(b_1,b_2)$ and $(c_1,c_2)$ outputs measurement outcomes (1,1) for any of the pairs. By correcting for both bit flips and phase flip errors in this manner, we can correct for any Pauli $X$,$Z$ and $Y$ errors, or any superpositions thereof. We note that the protocol above is also able to correct at least some 2 qubit errors, such as for instance simultaneous phase flips acting on different clusters, $Z_{a_{i}}Z_{b_{j}}$, or superpositions of a phase flip and a bit flip on 2 different qubits. For the full protocol, the encoding, decoding and correction processes are are entirely using concatenations of single qubit Pauli operations, 2 qubit CNOTs and CZs, and the coherence destroying measurements $K^0 = \ket{0}\bra{+}$ and $K^1 = \ket{1}\bra{-}$, all of which are incoherent operations.

The following is a concise list of the steps involved in the error correcting process:

\begin{enumerate}
	\item Initialise some quantum state $\ket{\psi} \coloneqq a \ket{0} + b \ket{1}$ whose information you want to protect, together with 2 copies of the classical state $\ket{0}$, and 6 copies of the state $\ket{+}$. This gives the initial state $\ket{\Psi_{ini}}_{abc} = \ket{+}_{a_1}\ket{0}_{a_2}\ket{+}_{a_3}\ket{+}_{b_1}\ket{\psi}_{b_2}\ket{+}_{b_3}\ket{+}_{c_1}\ket{0}_{c_2}\ket{+}_{c_3}$.
	\item Perform the first layer encoding unitary $U_1 \coloneqq U_{a_3a_2}U_{a_1a_2}U_{b_3b_2}U_{b_1b_2}U_{c_3c_2}U_{c_1c_2}$, such that $U_1\ket{\Psi_{ini}}_{abc} = \ket{0^{(++)}}_a \ket{\psi^{(++)}}_b \ket{0^{(++)}}_c$.
	\item Perform the second layer encoding $U_2 \coloneqq \cz{cb}\cz{ab}$ to get the final encoded state, such that $U_2 \ket{0^{(++)}}_a \ket{\psi^{(++)}}_b \ket{0^{(++)}}_c = a \ket{0_L}_{abc} + b \ket{1_L}_{abc} \coloneqq \ket{\Psi_L}$.
	\item Expose the state to some (single qubit) error model $\mathcal{E}$. Let the resulting density matrix be $\mathcal{E}(\ket{\Psi_L}\bra{\Psi_L})$.
	\item Decode the state by reversing the unitary, resulting in $\rho_{dec} = U_1^\dag U_2^\dag \mathcal{E}(\ket{\Psi_L}\bra{\Psi_L}) U_2 U_1$.
	\item Perform the measurement specified by the Kraus operators $K^0 = \ket{0}\bra{+}$ and $K^1 = \ket{1}\bra{-}$ corresponding to outcomes 0,1 on qubits $a_1, a_3, b_1, b_3, c_1, c_3$ . Perform classical projective measurements onto $\ket{0}\bra{0}$ and $\ket{1}\bra{1}$ on qubits $a_2, c_2$, recording the measurement outcomes.
	\item If any of the measurements of the qubit pairs $(a_1,a_3)$, $(b_1,b_3)$ and $(c_1,c_3)$ returns the measurement outcomes (1,1), you perform $Z_{b_2}$. If $(a_2, c_2)$ returns the outcome (1,1), you perform $X_{b_2}$. This measurement will destroy the coherence in the ancillary qubits.
	\item Tracing out every qubit except $b_2$ will return the original state $\ket{\psi}_{b_2}$.
	
\end{enumerate}

\section{Numerical Examples}

In this section, we provide numerical examples illustrating the effect of coherence on the error correction. We first initialize a pure quantum state $\ket{\psi(\theta)} = \cos(\theta/2)\ket{0} + \sin(\theta/2) \ket{1}$ parametrized by its polar angle on the Bloch sphere $\theta \in [0,\pi]$. This is the state whose quantum information we would like to protect from errors. We then initialize 2 ideal pure states in the classical state $\ket{0}$, together with 6 copies of the noisy state $(1-e)\ket{+}\bra{+} + e \openone/2$, parametrized by the noise parameter $e\in [0,1]$, before performing the full encoding process for the 9 qubit code described in the preceding section. $e = 0$ corresponds to the ideal case while $e=1$ corresponds to a completely noisy state. The noise parameter $e$ also serves as a proxy for the amount of coherence used to consumed during the error correction, starting from the maximal value of 6 qubits of coherence when $e=0$ and ending with no coherence at $e=1$. After the encoding, we perform a depolarizing channel on a random qubit which is our error model. The depolarizing channel performs the map $\Phi_{dep}(\rho) = (1-d)\rho + \frac{d}{3}(X \rho X + Y \rho Y + Z\rho Z)$, where $d \in [0,1]$. Finally, we perform the decoding and correcting process and plot the average fidelity of the state of the final qubit with the initial state $\ket{\psi}$. Note that the success  of the code against a depolarizing channel implies that it's capable of handling arbitrary single qubit errors arising due to either dissipative or non-dissipative interaction with the environment. The reason being that the Kraus operators corresponding to any such channel \cite{omkar2013} can be decomposed into $\openone,X,Y,Z$ operators. 

\begin{figure}[t]
\begin{subfigure}[t]{0.65\textwidth}
\centering
\includegraphics[width=1.0\linewidth]{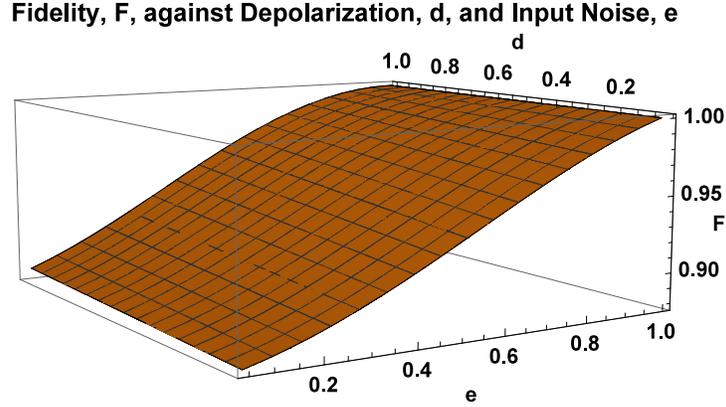} 
\caption{}
\end{subfigure}
\begin{subfigure}[t]{0.65\textwidth}
\centering
\includegraphics[width=0.9\linewidth]{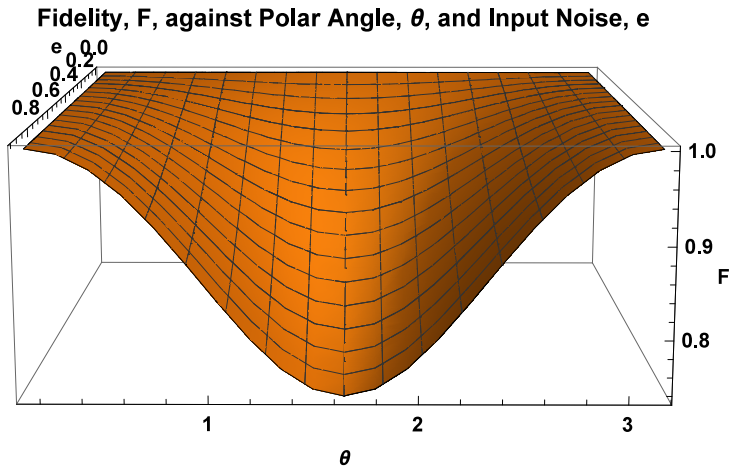} 
\caption{}
\end{subfigure}

\caption{3D plots of Fidelity against various parameters. (a) Fidelity against depolarization strength and input noise, for polar angle $\theta = \pi/4$. (b) Fidelity against input noise and the polar angle of the initial pure state at maximal depolarizing strength $d=1$.
}
\label{fig::fig1}
\end{figure}

From Figure~\ref{fig::fig1}(a), we can surmise that as a general trend, the fidelity decreases as the input noise increases, or alternatively, when the input coherence decreases. This decrease in fidelity is not uniform across all possible initial quantum states. From Figure~\ref{fig::fig1}(b), we see that the performance of the code varies depending on the polar angle of the initial state, being symmetric about the point $\theta = \pi/2$. This can be explained by the fact that the input noise, being a proxy for the amount of input coherence, is applied only to the non-classical resource. We have assumed that the incoherent qubit states $\ket{0}$ and $\ket{1}$ are classical and hence freely available. From this, we can surmise that the classical errors (bit flips) and the quantum errors (phase flips) is cleanly separated in our protocol. When the initial state is classical, i.e. when $\theta = 0, \pi$, the protocol only needs to correct for bit flips, which is a classical error, so no input coherence is necessary. On the other hand, when the initial state is quantum, especially in the region when $\theta = \pi/2$, we see that coherence is necessary to correct the error, only reaching maximum fidelity when the input coherence is maximal.

\begin{figure}[t]
\begin{subfigure}[t]{0.45\textwidth}
\centering
\includegraphics[width=1.0\linewidth]{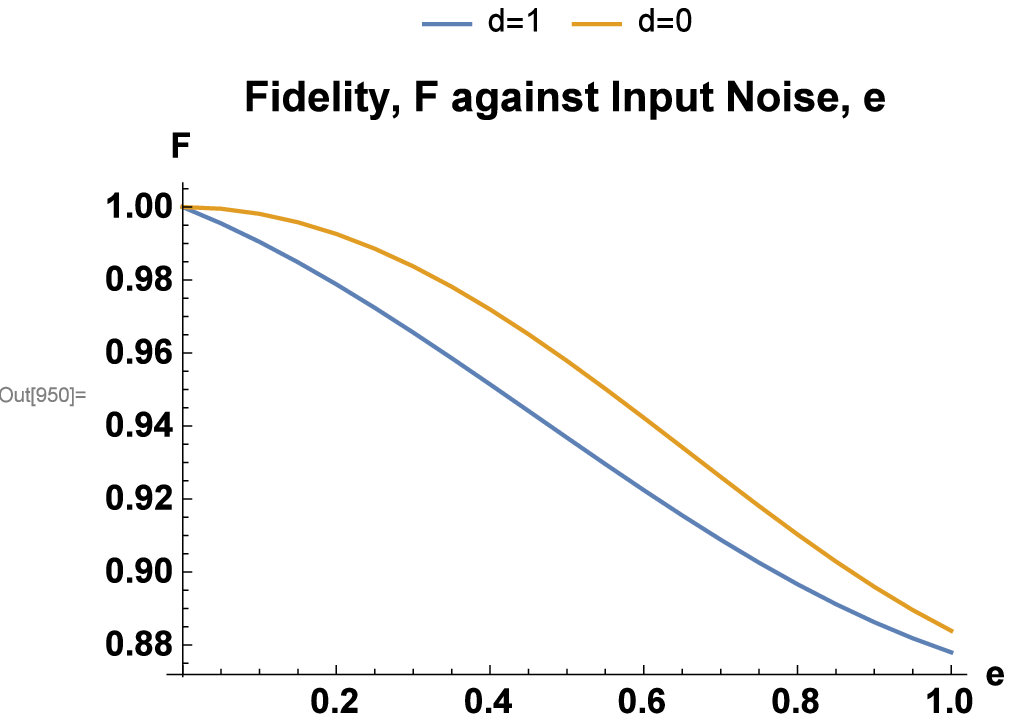} 
\caption{}
\end{subfigure}
\begin{subfigure}[t]{0.45\textwidth}
\centering
\includegraphics[width=1.0\linewidth]{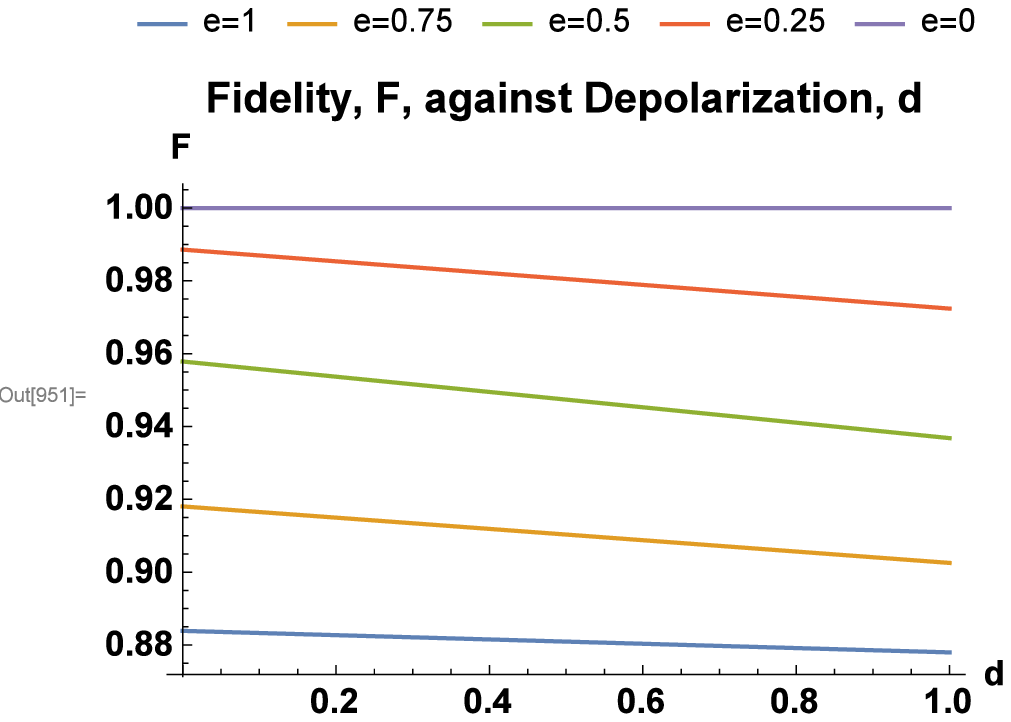}
\caption{}
\end{subfigure}
\begin{subfigure}[t]{0.45\textwidth}
\centering
\includegraphics[width=1.0\linewidth]{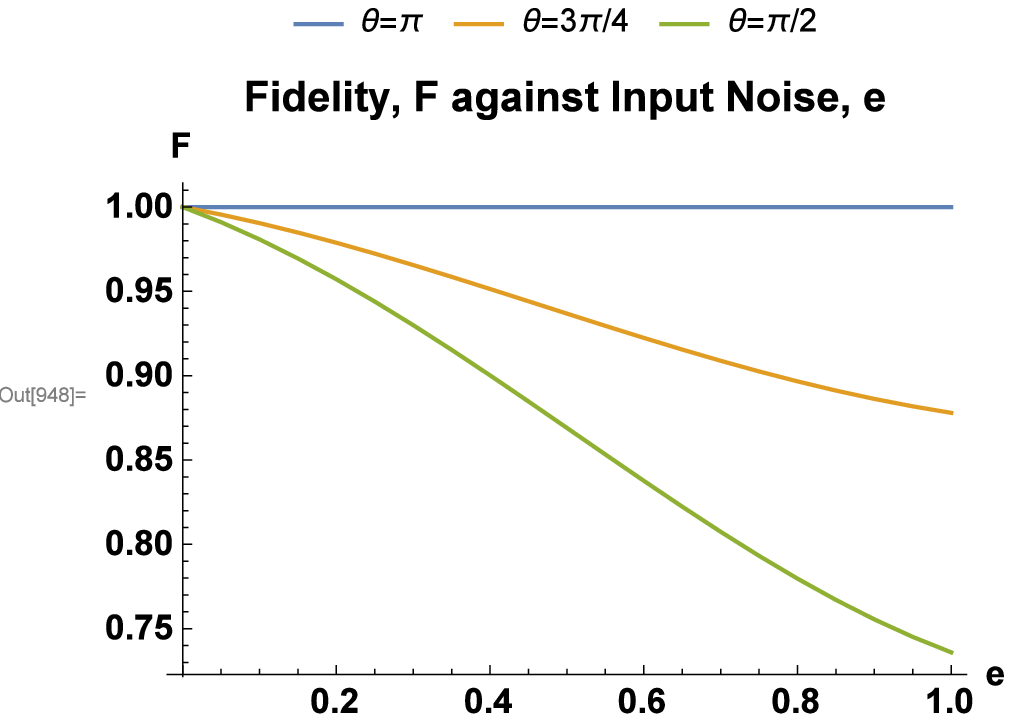}
\caption{}

\end{subfigure}

\caption{Plots of fidelity against various parameters. (a) Fidelity against input noise for various depolarization parameters. Polar angle is fixed at $\theta = \pi/4$. (b) Fidelity against depolarization strength, for various input noise. Polar angle is fixed at $\theta = \pi/4$. (c) Fidelity against input noise, for various polar angles of the initial state. Depolarization strength is maximal at $d=1$.
}
\label{fig::fig2}
\end{figure}

In Figure~\ref{fig::fig2}, we plot several vertical slices of the 3D plots, where again, we see that in the ideal case where the input noise $e=0$, the 9 qubit code is able to completely correct the single qubit depolarization, and hence any arbitrary single qubit error. From Figure~\ref{fig::fig2}(b), we observe that in the non ideal case when $e \neq 0$, fidelity is not 1 even when there is no depolarization (i.e. $d$ = 0 ). This is attributed to the fact that the the noise introduced when $e \neq 0$ can give rise to false error detections, thus introducing error to the encoded state when the correcting procedure is performed. In Figure~\ref{fig::fig2}(c), we see that the further away the initial input state is from $\ket{0}$ or $\ket{1}$, the lower the fidelity is and the more susceptible it is to $X$ and $Y$ errors. These errors will require coherence as input resources in order to correct.

\section{Conclusion}

In conclusion, we have studied an error correcting code which uses maximally coherent qubits as its initial quantum resource. We show that these codes are fully performed using only incoherent operations, thus both the resources at hand as well as the operations being performed fall under the regime identified by Baumgratz {\it et al.}~\cite{Baumgratz14}. We studied 3 different versions of this code, starting from a 2 qubit code conferring partial protection against a phase flip error, to a 3 qubit code providing complete protection against 1 phase flip error, and finally a 9 qubit code that is able to correct for any arbitrary single qubit error. The elementary quantum resources required for each of these codes are 1, 2 and 6 coherent qubits respectively. From this, we can see that even when the available quantum resource is a single coherent qubit, partial error correction can be achieved, while a pair will correct for a single quantum error, and 3 pairs can correct for arbitary single qubit errors, loosely corresponding to a pair of coherent qubits for each of the Pauli errors $X,Y$ and $Z$. The correction of the error will, however, require an (incoherent) syndrome measurement which will consume all the coherence used to perform the encoding and decoding process, thus necessitating additional coherent qubits in order to perform the protocol again. This is in line with the resource interpretation of quantum coherence. 

We do note that the protocol presented here is not necessarily optimal either in terms of the amount of input coherence or the number of ancillary qubits. Whether it is possible to reduce the resource requirements, in terms of the number of extra maximally coherent qubits required is an open question. However, it is known that the shortest code that can correct arbitary single qubit errors is a 5 qubit code~\cite{LaFlamme1996}, thus suggesting there is room for optimization if one were interested to reduce the amount of quantum resources being used as inputs. The success of the quantum error correction also depends on the nature of noise and hence the requisite amount of input coherence. For instance, one can construct codes utilizing less resources when the channel is known to be amplitude damping \cite{issac97}. We hope that the work presented here will spur further investigations into the role coherence plays in error correction and fault tolerant computations.

\section{Acknowledgements}

This work was supported by the National Research Foundation of Korea (NRF) through a grant funded by the Korea government (MSIP) (Grant No. 2010-0018295). K.C. Tan was supported by Korea Research Fellowship Program through the National Research Foundation of Korea (NRF) funded by the Ministry of Science, ICT and Future Planning (Grant No. 2016H1D3A1938100).

\end{document}